\documentclass[12pt,tightenlines,eqsecnum,floats,aps,nofootinbib,prd]{revtex4}
\usepackage{hyperref}
\usepackage{amsmath,amssymb,amsfonts,amsthm,amscd}
\usepackage{graphicx}
\usepackage{enumerate}
\usepackage{colordvi}
\usepackage{units}
\usepackage{epsfig}
\usepackage{natbib}
\usepackage{enumerate}
\usepackage{colordvi}
\usepackage{multirow}
\usepackage{afterpage}
\usepackage{pdflscape}
\usepackage{subfig}
\usepackage[utf8]{inputenc}

\def\be{\begin{equation}}
\def\ee{\end{equation}}
\def\ba{\begin{eqnarray}}
\def\ea{\end{eqnarray}}

\def\L{\mathcal{L}}
\def\H{\mathcal{H}}

\def\d{\mathrm{d}}

\def\V{\mathbf{V}}

\def\X{\mathbf{X}}
\def\R{\mathbb{R}}
\def\M{\mathcal{M}}
\def\c{\mathrm{c}}
\def\Vol{\mathrm{Vol}}
\def\det{\mathrm{det}}
\def\sign{\mathrm{sign}}

\begin{document}

\title{Scalar Fields and the FLRW Singularity}

\author{David Sloan}
\email{d.sloan@lancaster.ac.uk}
\affiliation{Lancaster University}

\begin{abstract}
\noindent The dynamics of multiple scalar fields on a flat FLRW spacetime can be described entirely as a relational system in terms of the matter alone. The matter dynamics is an autonomous system from which the geometrical dynamics can be inferred, and this autonomous system remains deterministic at the point corresponding to the singularity of the cosmology. We show the continuation of this system corresponds to a parity inversion at the singularity, and that the singularity itself is a surface on which the space-time manifold becomes non-orientable.

\end{abstract}

\maketitle

\section{Introduction}

Gravitational fields are not measured directly, but rather inferred from observations of matter that evolves under their effects. This is illustrated clearly by a gedankenexperiment in which two test particles are allowed to fall freely. In this the presence of a gravitational field is felt through the reduction of their relative separation. In cosmology we find ourselves in a similar situation; the expansion of the universe is not directly observed. It is found through the interaction between gravity and matter which causes the redshift of photons. The recent successes of the LIGO mission \cite{LIGO} in observing gravitational waves arise as a result of interferometry wherein the photons experience a changing geometry and when brought together interfere either constructively or destructively as a result of the differences in the spacetime that they experienced. What is key to this is the relational measurement of the photons; are they in or out of phase?

This relational behaviour informs our work. Here we will show how given simple matter fields in a cosmological setup, the dynamics of the system can be described entirely in relational terms. We see that this relational behaviour, being more directly related to physical observations, can be described without some of the structure of space-time. As such we treat general relativity as an operational theory; a means to the end of describing the relational dynamics of matter. This changes the ontological status of space-time. We do not treat the idea of scale in a four-dimensional pseudo-Riemannian geometry as absolutely fundamental to the description of physics, but rather as a tool through which dynamics can be calculated. This change of status is common to many approaches to fundamental physics; string theory \cite{Strings1,Strings2,Strings3,Strings4} often posits the existence of very small compact extra dimensions. In Loop Quantum Gravity \cite{LQG1,LQG2} the fundamental object is a spin network to which regular geometry is a an approximation at low curvatures, and in the cosmological sector this is responsible for removing the initial singularity \cite{LQC1,LQC2}. A minimalist approach is taken in Causal Set theory \cite{CST1,CST2} where the idea of geometry is rebuilt from causal relations between points, and a geometry is overlaid on top of these relations, and in Group Field Theory \cite{GFT} condensate states are interpreted as macroscopic geometries.

In this approach we will differ in one key way from the aforementioned approaches to fundamental theories of gravity. Rather than positing the existence of a more fundamental object on which our theory is based, we instead simply note that we do not have empirical access to the volume of the universe, and instead consider the relational evolution of observables. Aspects of this are captured in the Shape Dynamics \cite{Shapes1,Shapes2, Shapes3, Shapes4} program. At the heart of this is the idea that certain necessary factors in forming a space-time, such as the idea of an overall notion of scale, are not empirically measurable. As such, any choice of how dynamics is described in terms of these non-measurable quantities should not affect the evolution of measurable quantities. In previous work \cite{Through} we have shown that this leads to a unique continuation of Bianchi cosmologies (homogeneous, anisotropic solutions to Einstein's equations) through the initial singularity, and this has recently been extended by Mercati to include inflationary potentials for the matter \cite{FlavNew}. There are two principal reasons why this is possible; the first is that the system exhibits ``Dynamical Similarity" \cite{DynSim} and thus solutions can be evolved in terms of a smaller set of variables than are required to describe the full phase-space. The second is that the equations of motion for these variables are Lipschitz continuous even at the initial singularity. By the Picard-Lindel\"of theorem, they can be uniquely continued beyond this point and reveal that there is a qualitatively similar, but quantitatively distinct, solution on the other side. In this paper we will show that the same results hold when working with scalar fields in a flat Friedmann-Lema\^itre-Roberston-Walker (FLRW) cosmology. 

This paper is laid out as follows. In section \ref{FLRWSec} we recap the dynamics of flat FLRW cosmologies in the presence of scalar fields, and express the dynamics as a flow on the usual phase space. Then in section \ref{DynSimSec} we show the role of dynamical similarity in these spacetimes, establishing a vector field on phase-space whose integral curves take solutions to those which are are indistinguishable. This allows us to formulate the more compact description of the system which is well-defined at and beyond the singularity. We then show some general features of such systems. In section \ref{FreeFieldsSec} we show how massless noninteracting scalar fields provide this continuation, and in section \ref{ShapeSec} we show the fully intrinsic form of the equations of motion when interactions are reintroduced. We examine how one can reconstruct a geometrical interpretation on the other side of the singularity in section \ref{NonOrientableSec} and show that this would appear to be an orientation flip when viewed in this way. Finally in section \ref{BeyondSec} we show how the results we have obtained extend beyond the isotropic case and make contact with prior results, and give some concluding thoughts in section \ref{SecDiscussion}. 

\section{FLRW Cosmology with Scalar Fields} \label{FLRWSec}

We will examine the dynamics of scalar fields in a flat FLRW cosmology. To retain the homogeneity and isotropy of our solutions, we will assume the same holds for our scalar fields, and thus each field has only temporal variation. The metric takes the form:
\be \d s^2 = -\d t^2 + a(t)^2 \left(\d x^2+\d y^2+\d z^2\right) \ee
It is important to note that there are two tetrad representations of this system which are compatible with the geometry corresponding to left-handed and right-handed orientations $g=\eta(\mathbf{e},\mathbf{e})$, with the choices $\mathbf{e}_L = (\d t,a\d x,a\d y,a\d z)$ and $\mathbf{e}_R = (dt,-a\d x,-a\d y,-a\d z)$. In fact, since the form $\eta$ is bilinear, we could have chosen to distribute the - signs with any of these components, however since we will be primarily interested in the behaviour of the spatial parts across the initial singularity, we choose to keep a time direction fixed and will only be interested in the relative signs of the one-forms across $t=0$. 

Dynamics are derived from the Einstein-Hilbert action for gravity minimally coupled to matter, which has the usual scalar field Lagrangian. Our spacetime is topologically $\R \times \Sigma$ where the spatial slice $\Sigma$ can be $\R^3$ or $\mathbb{T}^3$. In the case of $\R^3$ we choose a fiducial cell to capture the entire system since homogeneity means that the dynamics of the entire space can be determined by the dynamics of any chosen subregion, and thus we avoid infinities.
\be S = \int \sqrt{g}(R - \mathcal{L}_m) =  \int_\R \int_\Sigma a^3\left(6(\frac{\ddot{a}}{a} + \frac{\dot{a}^2}{a^2}) - \frac{\dot{\phi}^2}{2} + V(\phi) \right) \ee
In order to simplify the algebra, in the following we make the choice to work with the volume instead of the scale factor, $v=a^3$. The momentum conjugate to $v$ is proportional to the Hubble parameter, $h=p_v=\frac{4\dot{v}}{v}$ and that to $\phi$ is $p_i=v\dot{\phi}_i$, where we choose to denote this conjugate momentum $h$ to avoid notation clashes. The Hamiltonian and symplectic structure are given:
\be \H = v\left(-\frac{3h^2}{8} + \frac{\vec{p}^2}{2v^2} + V(\vec{\phi})  \right) \quad \quad \omega = \d h\wedge \d v + \d\vec{p} \wedge \d\vec{\phi}\ee
The Hamiltonian vector field, $\X_\H$ describes the evolution of a solution in phase-space. It is determined uniquely through the global invertibility of the symplectic form (summing over repeated indices of the scalar field and its momentum):
\be \d\H = \iota_{\X_\H} \omega \rightarrow \X_\H =  \frac{3 v h}{4} \frac{\partial}{\partial v} - \frac{p_i^2}{v^2} \frac{\partial}{\partial h} +\frac{p_i}{v} \frac{\partial}{\partial \phi_i} - v\frac{\partial V}{\partial \phi_i} \frac{\partial}{\partial p_i}  \label{HVF} \ee
The dynamics of the matter present is given by the Klein-Gordon equation, which corresponds to the usual Hamiltonian dynamics of the scalar fields given the above:
\be \label{KG} \ddot{\phi_i} + \frac{3h}{4} \dot{\phi_i} + \frac{\partial V}{\partial{\phi_i}} = 0 \ee
and the dynamics of the geometry is given by the Friedmann equation:
\be h^2 = \frac{8}{3}\left(\frac{\dot{\phi}^2}{2}+V(\phi)\right) \ee
In the case where there is no potential for the scalar field, we can solve these analytically to see $v=v_o t$, $\phi_i = A_i \log t + B_i$. 

The singularity of this system corresponds to the fact that along its orbit on phase space, $\X_\H$ reaches a point at which it is no longer integrable. From the Picard-Lindel\"of theorem, this arises because uniqueness of solutions to the equations of motion fails when coefficients of the basis vectors ($\frac{\partial}{\partial h}$ etc) are not Lipschitz continuous. We see that this can occur in two ways; the first is that some of the phase space variables will tend to infinity. We will show that this can be solved through a compactification which is brought about by considering only relational variables. The second point at which Lipschitz continuity can fail is when $v=0$. However, it turns out that this system can be expressed in such a way that $v$ is not strictly required for evolution. This happens because the Hamiltonian has solutions which are dynamically similar, and thus the evolution can be described on a contact manifold in terms of relational variables.

\section{Dynamical Similarity in FLRW Cosmology} \label{DynSimSec}

Since the total Hamiltonian $\H=0$ is a constraint, we can use this to replace $h$ in equation (\ref{KG}) and express the dynamics of the matter fields purely in a closed form.
\be \label{KG2} \ddot{\phi_i} + \dot{\phi_i} \sqrt{24 V(\phi) + 12 \dot{\vec{\phi}}^2} + \frac{\partial V}{\partial{\phi_i}} = 0 \ee
On first glance this may appear surprising; we began with a $2n+2$ dimensional phase-space subject to a single constraint, and have arrived at a set of $n$ coupled differential equations which are second order in the fields alone. However, the reason for this reduction is that there is a somewhat hidden symmetry of this system; it exhibits dynamical similarity \cite{DynSim} under rescaling both the momenta of the matter fields and the volume (measured through to an appropriately chosen fiducial cell) of the spatial slice. The dynamical similarity that arises is an exact consequence of the fact that the dynamics of the system chosen should not be affected by the choice of a cell by which it is measured. 

In this system we have universal coordinates on phase space and thus we can define a non standard canonical transformation through the vector field 
\be \V = v \frac {\partial}{\partial v} + \vec{p_\phi} \frac{\partial}{\partial \vec{p_\phi}} \ee
Following the procedure laid out in \cite{DynSim}, we see that this system has an autonomous subsystem of dynamics of the invariants of $\V$, given by the contact Hamiltonian $H^c$ and contact form $\eta$
\be \H^{\c} = -\frac{3h^2}{8} + \frac{\dot{\vec{\phi}}^2}{2} + V(\vec{\phi}) \quad \quad  \eta = -\d h + \dot{\vec{\phi}} \d \vec{\phi} \ee
It is thus clear that this is a system of $2n+1$ invariants of $\V$ subject to a single constraint, and thus the fact that we can describe the relational motion in terms of just the matter fields themselves is no longer surprising. The dynamics of contact systems is an interesting and active area of study \cite{Bravetti,Leon}. The contact space is odd-dimensional, and thus differs from symplectic dynamics. The is one particular difference which is important in the cosmological case: contact systems are `frictional'. It is clear from the contact form that one of the variables is distinct from the rest, as it appears as a coordinate without a corresponding momentum. Thus there exists a vector field $\nu$ such that $\nu(\eta)=1, \nu(\d\eta)=0$. This is called  the `Reeb vector field', which in our case is $\nu=\frac{\partial}{\partial h}$, through which friction becomes apparent. The contact Hamiltonian vector field $\X_{\H^{\c}}$ is determined uniquely through
\be \iota_{\X_{\H^{\c}}} \d\eta + \eta(\X_{\H^{\c}}) \eta =\d\H^{\c} - (\nu(\H^{\c})+\H^{\c})\eta \ee
Thus our system has a contact Hamiltonian vector field given by
\be \X_{\H^{\c}} = -\dot{\phi}_i^2 \frac{\partial}{\partial h} + \dot{\phi}_i\frac{\partial}{\partial \phi_i} - \left(\frac{\partial V}{\partial \phi_i} +\frac{3h\dot{\phi_i}}{4}\right) \frac{\partial}{\partial \dot{\phi}_i}  \ee
We see immediately that we have overcome one of the failures of Lipschitz continuity; since this system does not refer to $v$ at any point, its vanishing does not impede the integration of the contact Hamiltonian vector field. Thus the points at which we may encounter problems are where one or more of $\phi,\dot{\phi},h$ become infinite. However as we have previously noted, this is solved by the compactification induced in choosing relational variables.

Since the volume $v$ is not a part of the relational system itself, and is never needed to find the equations of motion of our contact system, we will not consider it to be fundamental to the description of our system. Rather we will note that we can completely integrate our equations of motion to find the behaviour of $h$ for all times. We then create a geometrical model which corresponds to our system by choosing a value for $v$ at any given time and finding its time evolution by quadrature. In doing so we can reconstruct a space-time geometry which exactly reproduces the dynamics of the FLRW system. However since $v$ is not fundamental to our description, neither is its role as volume in this system; the singularity at $v=0$ is thus not a problem; we have simply reached the boundary of the set of solutions that can be reproduced using our chosen space-time geometry. As we will see this is still a regular point in the relational description, and thus we can create another space-time geometry beyond $v=0$ corresponding to the evolution of the relational system past this point. The singularity of GR is thus a failure of the description of physics in terms of an orientable manifold.  

The frictional nature of a contact system is manifested in two ways; the first is that the contact Hamiltonian, $\H^{\c}$ is not necessarily preserved along its integral curves:
\be \frac{\d \H^{\c}}{\d t} = -\H^\c \frac{\partial \H^{\c}}{\partial h} \ee
however, we note that as this is a constraint in cosmology, $\H^{\c}=0$, and thus the system is conserved. The second is that Liouville's theorem is modified for the system \cite{Bravetti2}. As the space is odd dimensional, we cannot make a volume form in the usual way by taking exterior products of the symplectic form. Instead, the canonical volume form on a contact space is given 
\be \Vol = \eta \wedge \d\eta^n \ee
By evaluating the time derivative of this form, we see that the Hamiltonian flow on contact space has a divergence, and thus 
\be \dot{\Vol} = -(n+1) \frac{\partial \H^{\c}}{\partial h} \Vol \ee
which gives rise to attractors on the space of invariants. Since the Reeb vector field in our cosmological systems is along the direction of the Hubble parameter, its dual in the original symplectic system is the expansion volume; $\iota_\frac{\partial}{\partial h} \omega = \d v$. Thereby we see that the friction present in the matter systems arises as a consequence of the expansion of the universe.  

\section{Free Fields} \label{FreeFieldsSec} 
\label{Free}
For clarity we will focus on a system with two fields present $\vec{\phi} = (\phi_1, \phi_2)$, though the procedure described is entirely general. A theorem due to Foster \cite{Foster} shows that on approach to a singularity in such a spacetime for a broad class of potentials the scalar fields asymptote to their massless ($V=0$) behaviour. Therefore I shall first work in the case of free fields. As such the dynamics are easy to express:
\be \frac{\d h}{\d t}=-\frac{3h^2}{4} \rightarrow h=\frac{4}{3t} \quad \frac{\d\dot{\phi_i}}{\d t} = \frac{3h\dot{\phi}}{4} \rightarrow \dot{\phi_i} = \frac{A_i}{t}  \ee
Thus we see solutions in parameter time given by
\be \phi_1 = A_1 \log t + B_1 \quad \phi_2 = A_2 \log t + B_2 \ee
If $t$ is not a direct observable then we have our complete motion in the $(\phi_1,\phi_2)$ plane determined along the line $\phi_2 = \frac{A_2}{A_1} \phi_1 + \frac{A_1 B_2 - A_2 B_2}{A_1}$. This is simply the equation of a straight line. Note that there the solution is degenerate in choices of $A_i$ and $B_i$ since these four variables determine a single straight line in $\mathbf{R}^2$. This is unsurprising as the physics of the underlying system is unchanged under a shift of $B_1$ and $B_2$ -- these leave $\dot{\phi_i}$ unchanged, and similarly we could reparametrize the time direction changing $A_1$ and $A_2$ but retaining their proportion. 

We can express the system on shape space by making the compactification of the system on a sphere:
\begin{equation}\label{SphericalShapeCoordinates}
\left( \begin{array}{c}\phi_1\\\phi_2\end{array}\right) =|\tan\beta| \left( \begin{array}{c}
\cos\alpha \\ \sin \alpha
\end{array} \right)
\end{equation}
This compactification can be visualized by considering taking a unit sphere tangent to the $(\phi_1 - \phi_2)$ plane touching at the north pole. The mapping takes a point $P$ on the plane onto the sphere by passing a line from the center of the sphere to $P$ and identifying this with the point of intersection with the sphere. 

Thus we have kinematical equation for $\alpha$ and $\beta$:
\ba \dot{\beta} &=&  \cos^2\beta \left(\dot{\phi_1} \cos\alpha +\dot{\phi_2} \sin\alpha \right) \\
      \dot{\alpha} &=& \cot\beta \left(\dot{\phi_2} \cos\alpha - \dot{\phi_1} \sin\alpha \right) \ea
Therefore we can find the relational equation on a solution to the equations of motion:
\be \frac{\d\alpha}{\d\beta} = \frac{1}{\sin^2\beta} \frac{A_2 B_1 - A_1 B_2}{A_1 \cos\alpha + A_2 \sin\alpha} \ee
This equation appears singular at the point where $\tan\alpha = -A_1/A_2$, however note that this is simply the point of closest approach of the line to the origin in the $(\phi_1,\phi_2)$ plane, and therefore would be a stationary point of $\beta$ on a solution. Which we can integrate to find:
\be A_1 \sin\alpha - A_2 \cos\alpha = (A_1 B_2 - A_2 B_1) \cot\beta \ee
which is the equation of a geodesic on the sphere. The constant of integration is set to zero to match the solution we are representing here with the representation of the scalar fields in terms of time. The equator of the sphere is at $\beta=\pi/2$ which represents the singularity in the FLRW system, but is again a regular point in the relational shape system. We can thus recover the space that is covered by our solutions in the $(\phi_1,\phi_2)$ plane by noting that our solutions are parametrized by $\alpha$, which is monotonic in time. Hence
\begin{equation}\label{FreeShapeSolutiononPlane}
\left( \begin{array}{c}\phi_1\\\phi_2\end{array}\right) =\left|\frac{A_1 B_2-A_2 B_1}{A_1 \sin\alpha - A_2 \cos\alpha} \right| \left( \begin{array}{c}
\cos\alpha \\ \sin \alpha
\end{array} \right)
\end{equation}
It is obvious that this solution is periodic in $\alpha$ with period $2\pi$. We thus see that the complete solution when carried beyond the initial singularity of the FLRW spacetime continues in the $(\phi_1,\phi_2)$ plane along a line parallel to the original solution but with opposite impact parameter, and travelling in the opposite direction in parameter time. Thus the extended system is given $\phi_i =  (A_i \log |t| +\sign(t) B_i)$, valid wherever $t \neq 0$. 

It is interesting to note here that it is only when we introduce potentials that the functional form of the gravitational action comes into play; if we had replaced the $h^2$ term with $f(h)$ in the contact Hamiltonian, $\H^{\c}$ then $\frac{\d\phi_1}{\d\phi_2}$ would be unaffected, and we'd still have the space of straight lines, and so the shape space compactification would have been the same. This is broken by interactions between the fields, as we will see in section \ref{ShapeSec}.

Let us now consider a set of $n$ scalar fields. Each of these will have have its dynamics given in terms of time $t$ by 
\be \phi_i = A_i \log t  + B_i \ee
for constants $A_i$ and $B_i$. Following the same procedure as above, we expect that there will be a well defined relational motion in terms of the compactified shape space. We begin by using coordinates on the $n$-sphere:
\begin{equation}\label{SphericalShapeCoordinates2}
\left( \begin{array}{c}\phi_1\\ \phi_2 \\...\\ \phi_n\end{array}\right) =|\tan\beta| \left( \begin{array}{c}
\cos\alpha_1 \\ \sin \alpha_1 \cos \alpha_2 \\ ... \\  \sin \alpha_1 ... \sin \alpha_{n-1}
\end{array} \right)
\end{equation}
The geometric idea here is the same as above -- considering the hyperplane in the $\phi_i$ and mapping onto an $n$-sphere by placing the sphere tangential to the plane at the north pole, and associating points on each via a line from the center of the sphere to the plane.

The radial component of our motion is contained in $\tan \beta$. We know that for free fields this radius will tend to infinity as we approach the singularity at $t \rightarrow 0$, which in these terms is the equator of the $n$-sphere, $\beta=\pi/2$.Hhence we again parametrize our motion in terms of $\beta$ and examine the dynamics as we approach this point. From the equations of motion for the scalar fields we can make the relation between $\beta$ and $t$ explicit:
\be \tan^2 \beta = \sum \phi_i^2 = \sum A_i^2 (\log t)^2 + A_i B_i \log t + B_i^2 \ee
and hence to leading order $|\tan \beta|$ approaches infinity as $\log t$. From this we see that the leading order contribution to $\dot{\beta}$ is given by $1/t(\log t)^2$. 

For the $\alpha_i$ we can perform a similar analysis of their asymptotic structures. We first note that 
\be \alpha_i = \frac{\sqrt{\phi_{i+1}^2 + ... + \phi_n^2}}{\phi_i} \ee
and we find its velocity to be
\be \dot{\alpha_i} = \frac{\cos^2 \alpha_i}{\phi_i^2 \sqrt{\phi_{i+1}^2 + ... + \phi_n^2}} \left(\phi_i(\phi_{i+1}\dot{\phi}_{i+1} + ... +  \phi_n \dot{\phi}_n) - \dot{\phi}_i (\phi_{i+1}^2 + ... + \phi_n^2 \right) \ee
Here we note that the bracketed term on the right hand side appears to have a term that grows as $\log(t)^2/t^2$. However this term is exactly zero, and hence the leading order contribution is in fact 
\be \dot{\alpha}_i = \frac{\cos^3 \alpha_i}{\sin^3 \alpha_1 ...\sin^3 \alpha_{i-1} \tan^3 \beta \sin \alpha_i} \left(  (A_i \prod_{j>i} A_j B_j - B_i \prod_{j>i} A_j^2) \frac{\log(t)}{t} + o(t^{-1}) \right) \ee 
and therefore to leading order $\dot{\alpha}$ approaches the singularity as $1/(t\log t)^2$. Hence at the singularity we find that $\frac{d\alpha_i}{d\beta}$ tends to a constant and the system can be continued beyond this point. It is a (long) exercise to show that the solutions of this system correspond to geodesic motion on the $n$-sphere. 
 
\section{Full Shape Dynamical System} \label{ShapeSec} 

Since we know what the asymptotic behaviour will look like for most well-behaved potentials,\footnote{These will turn out to be potentials $V$ where $e^{-\sqrt{\frac{3}{4}(\phi_1^2+\phi_2^2)}}V$ and its derivatives $e^{-\sqrt{\frac{3}{4}(\phi_1^2+\phi_2^2)}}\frac{\partial V}{\partial \phi_i}$ are finite.} we can set up our complete shape system. Initially it might appear that a good choice would be to pick $p=\sqrt{\dot{\phi}_1^2+\dot{\phi}_2^2}$ and $\tan{\theta}=\frac{\dot{\phi}_2}{\dot{\phi}_1}$. However, as the system asymptotes to the motion of a free field, the velocity of the field will become parallel to the field ($\theta \rightarrow \alpha$), therefore we cannot simply break the velocity into polar coordinates. Likewise asymptotically the velocity of the field will outgrow the field values ($p \gg \tan{\beta}$). It turns out that a good set of variables are
\be \lambda = \frac{\sin(\alpha-\theta)}{\cos\beta} \quad \chi = p e^{-\frac{\sqrt{3}}{2}|\tan\beta|} \label{defs} \ee
which are asymptotically distinct from $\alpha$ and $\beta$ and thus contain solution determining data, and are Lipschitz continuous at $\beta \rightarrow \frac{\pi}{2}$. The complete dynamics of our system can now be expressed completely in these terms. Denoting a derivative with respect to $\beta$ with a prime, 
\ba \label{EoM} \alpha' &=& \frac{\lambda}{\sin{\beta}\sqrt{1-\lambda^2 \cos^2\beta} } \nonumber \\
      \lambda' &=&  \frac{\lambda}{\cos{\beta}} \left(\sin{\beta} - \frac{1}{\sin\beta}\right) +\frac{V_2 \cos(\theta) - V_1 \sin(\theta)}{\chi^2 e^{\sqrt{3} |\tan\beta|} \cos^2\beta}  \nonumber \\
      \chi' &=& \frac{\sqrt{3} \chi}{2\cos^2\beta\sqrt{1-\lambda^2 \cos^2{\beta}}} \left(\sqrt{1+\frac{2V}{\chi} e^{-\sqrt{3}|\tan\beta|}} - \sqrt{1-\lambda^2 \cos^2{\beta}} \right) \nonumber \\ 
              &-& \frac{\cos(\theta)V_1 + \sin(\theta)V_2}{\cos^2\beta \sqrt{1-\lambda^2 \cos\beta^2} \chi^3 e^{\frac{3\sqrt{3}}{2}|\tan\beta|}}  \ea
Thus our system is asymptotically (assuming that $V$ is well behaved) $\alpha' = \lambda, \lambda' = 0, \chi' = -\frac{\sqrt{3} \lambda^2 \chi}{4}$. The equations of motion (\ref{EoM}) are Lipschitz continuous and thus we satisfy the conditions of the Picard-Lindel\"of theorem. This holds even at $\beta=\pi/2$, which corresponds to the initial singularity of the cosmological system. Therefore there is a unique continuation of any given solution through $\beta=\pi/2$.

As we have shown that the system of equations described in equations (\ref{EoM}) is well defined at the initial singularity, it is logical to then ask if there are places at which this description breaks down. It is clear from the evolution of $\alpha$ that the equations are singular when $\lambda \cos\beta=\pm 1$. From the definition of $\lambda$ we see that this is a point at which the velocity of the scalar fields is orthogonal to their position. Since $\tan\beta$ represents the distance from the origin in the $(\phi_1,\phi_2)$ plane, this is unsurprising. At this point the motion of the fields has no radial component, and hence $\beta$ is unchanging while the other variables are, and hence derivatives of these with respect to $\beta$ will become infinite. The motion remains well defined when expressed in terms of the fields and their velocities alone, however, and thus this is simply a poor choice of representation of the system at this point. It is entirely analogous to the singularity in $\frac{\d r}{\d\theta}$ in the polar coordinate description of a straight line; there at the point of closest approach to the origin $r$ is it a minimum whilst $\theta$ changes, and thus in these variables the system appears singular. However, the equation $y=mx+c$ remains well defined at this point. The question then arises as to whether this could be the asymptotic behaviour as $\beta \rightarrow \pi/2$. The answer is that if $V$ obeys the conditions of regularity then it cannot, as the motion asymptotes to that of a free field.

Recovering the free field in these circumstances amounts to setting $V$ and its derivatives to zero. Upon doing so we note two interesting features; the first is that $\chi$ becomes unimporant in dynamics, as it only enters the equations of motion for $\alpha$ and $\lambda$ through terms proportional to derivatives of the potential. Hence we can integrate the equation of motion for $\lambda$ directly in this case to obtain
\be \lambda_\textrm{free} = \frac{\lambda_\textrm{0}}{\sin\beta} \label{lambdafree} \ee
wherein $\lambda_0$ is a constant. Reintroducing this into the equation of motion for $\alpha$ gives the equation for a straight line in the $(\phi_1,\phi_2)$ plane.  Second, we see that this feature only relied upon there being non-zero derivatives of $V$. Had we introduced a pure cosmological constant term (corresponding to $V$ being constant) the equations of motion would have been unaffected. From equation \ref{lambdafree} we see that 
\be \lambda \cos\beta = \frac{\lambda_\textrm{0}}{\tan{\beta}} \ee
and so is decreasing (to zero) at the singularity. Thus when the fall-off conditions for the potential are obeyed, the system is always integrable through the singularity. 

Extending our system to $n$ scalar fields is a simple exercise; first we note that at any point in the evolution of the system we can choose coordinates for the scalar fields such that the position and velocity of the fields are in the plane spanned by $\phi_1$ and $\phi_2$. If we describe the positions and velocities of our fields in terms of spherical coordinates, with positions given by equation \ref{SphericalShapeCoordinates2} and velocities given by
\begin{equation}
\label{SphericalShapeCoordinates3}
\left( \begin{array}{c}\phi_1\\ \phi_2 \\...\\ \phi_n\end{array}\right) =p \left( \begin{array}{c}
\cos\theta_1 \\ \sin \theta_1 \cos \theta_2 \\ ... \\  \sin \theta_1 ... \sin \theta_{n-1}
\end{array} \right)
\end{equation}
then this amounts to setting $\alpha_2,...,\alpha_{n-1}$ and $\theta_2,...,\theta_n$ to zero locally. Thus the equations of motion for the position and velocity in the plane remain as in equations \ref{EoM} locally, defining $\lambda=\frac{\sin(\alpha_1-\theta_1)}{\cos\beta}$;
\ba \label{EoM2} \alpha_1' &=& \frac{\lambda}{\sin{\beta}\sqrt{1-\lambda^2 \cos^2\beta} } \nonumber \\
      \lambda' &=&  \frac{\lambda}{\cos{\beta}} \left(\sin{\beta} - \frac{1}{\sin\beta}\right) +\frac{V_2 \cos(\theta_1) - V_1 \sin(\theta_1)}{\chi^2 e^{\sqrt{3} |\tan\beta|} \cos^2\beta}  \nonumber \\
\ea
The presence of the potentials causes the motion to accelerate from the plane 
\be \theta'_j = \frac{V_{j+1}}{\chi^2 e^{\sqrt{3} |\tan\beta|} \cos^2\beta \sqrt{1-\lambda^2 \cos^2 \beta} } \ee
and hence again we see that for potentials which are well behaved (in the sense that both the potential and its derivatives fall off sufficiently) the motion becomes asymptotically planar and again describes a geodesic on the $(n-1)$-sphere. Similarly, we see that if the potentials are well behaved then the equations remain Lipschitz continuous, and hence the Picard-Lindel\"of theorem guarantees that there is a unique trajectory which continues the solution through $\beta=\pi/2$.

\section{Conserved Quantities on Non-orientable Manifolds} \label{NonOrientableSec}

The singularity of FLRW cosmology is a point at which the spacetime geometry endows (any chosen fiducial cell embedded in) the spatial slice with zero volume. The volume form, induced by the space-time volume form pulled back to a $t=constant$ slice, is $\Vol_\Sigma = v \d x\d y\d z$. The existence of a volume form is equivalent to orientability of the manifold itself; non-orientable manifolds of dimension $d$ do not have any nowhere-vanishing d-forms. In extensions of our system beyond a point at which the volume form vanishes, it is therefore natural to ask whether this might be an indication that the manifold is not globally orientable. The volume form is directly linked to the conserved quantities of any Lagrangian system through Noether's theorem; it is almost universally assumed that any manifold providing a basis for physics must be orientable. Here we will relax this somewhat.

To provide an example of such a system, let us consider a Mobius strip formed by taking a section of $\R^2$ and placing the usual twist on the boundaries. Specifically we will take
\be \M = \frac{(0,1) \times [0,1]}{\sim} \quad (0,y) \sim (1,1-y) \ee
and we will endow this space with a metric $\d s^2 = \d x^2+\d y^2$ everywhere except the points of identification $x=0,1$. This metric admits two diads up to a choice of overall sign; $\mathbf{e}_1 = \d x, \d y$ and $\mathbf{e}_2 = \d x, -\d y$. Hence our choices of volume form are $\Vol=\pm \d x\wedge \d y$. Let us consider the motion of a free particle on this manifold. The Lagrangian for such a particle (again, defined away from the edges) is 
\be \L = \frac{\dot{x}^2}{2} + \frac{\dot{y}^2}{2} \ee
And we find from this that the two momenta $P_x=\dot{x}, P_y=\dot{y}$ are conserved quantities. This is, of course, the statement that free particles follow the geodesics of the manifold. In the Hamiltonian language we know that the system is determined by a Hamiltonian and symplectic structure
\be \H = \frac{P_x^2}{2}+\frac{P_y^2}{2} \quad \omega = \d P_x \wedge \d x + \d P_y \wedge \d y \ee

\begin{figure}[h]
\includegraphics[height=100pt]{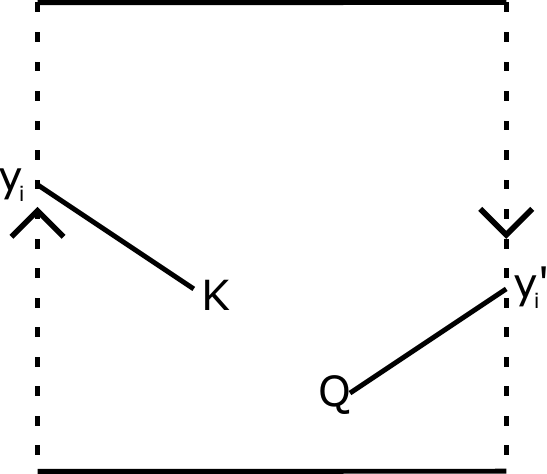} \quad\quad\quad\quad\quad\quad
\includegraphics[height=100pt]{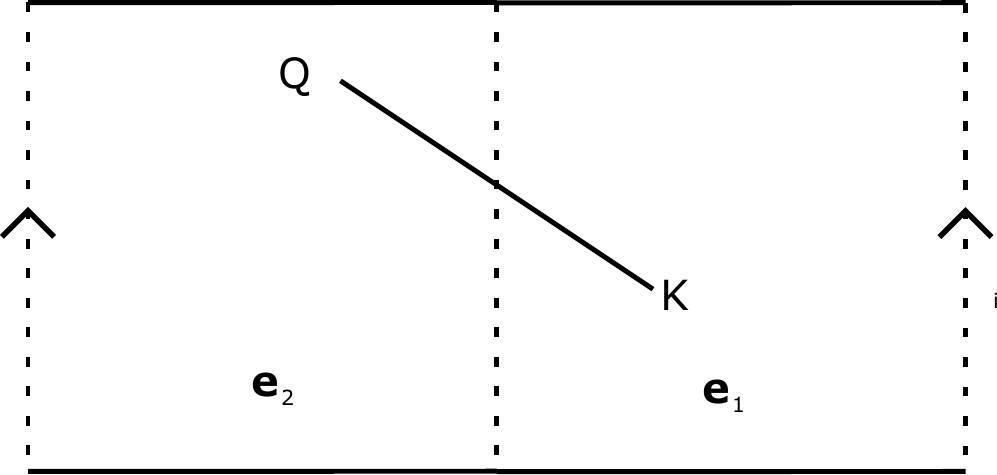}
\caption{The geodesic on the Mobius strip shown on the left on a single copy of the space. We see that the line is discontinuous under the identification and has a sharp corner at the join. On the right is the same line but represented in a double cover of the space where we have used the diad $\mathbf{e}_2$ on the left and $\mathbf{e}_1$ on the right, swapping the orientation between copies of the space. With this choice we recover what we would expect a geodesic to look like on the double cover with metric given in terms of $x$ and $Y$.}
\label{GeoMobius}
\end{figure}

Suppose we now want to extend this to what happens to a particle that crosses $x=0$, say. In such a case we can extend geodesics across this point by insisting that the path taken is a minimum of the distance as defined on all points away from $x=0$ and that the path is continuous across the join. Away from the join, the geodesics will be geodesics of the metric (straight lines in $\R^2$) and so we need only consider the point of intersection with the join, and connect the points with straight lines away from it. On doing so we find that geodesics connecting two points $K=(K_x,K_y)$ and $Q=(Q_x,Q_y)$ are given by minimizing
\be \Delta(K,Q) = \Delta(K,(0,y_i))+\Delta(1,(1-y_i),Q)= \sqrt{K_x^2+(K_y-y_i)^2} + \sqrt{(Q_x-1)^2+(Q_y-(1-y_i))^2} \ee
across all choices of the $y_i$, the coordinate of the intersection of the path with $x=0$. Upon doing so we find that the geodesic crossing the join between $K$ and $Q$ is equivalent to taking a second copy of the space attached at the join but flipped in the $y$ direction, and finding a geodesic on the combined space with the metric $\d s^2 = \d x^2 + \d Y^2$, where $Y=y$ for $x>0$ and $Y=1-y$ for $x<0$. This is equivalent to saying that we match across the join by swapping $\mathbf{e}_1$ for $\mathbf{e}_2$. This is shown in figure \ref{GeoMobius}. One useful way to visualize this is to consider a physical M\"obius strip formed by taking a strip of paper and connecting it with the usual twist. The double cover amounts to viewing the two sides of the paper, and our geodesic is a straight line along the paper which starts on one side and ends on the other. A circle with an arrow pointing clockwise on one side will appear to point anticlockwise when viewed from the other side, illustrating the choice between $\mathbf{e}_1$ and $\mathbf{e}_2$.

If we turn our attention back to the conserved quantities it would appear that across the join a particle following a geodesic will have preserved $P_x$ but $P_y \rightarrow -P_y$. There are several ways to interpret this. We could first consider that what is actually conserved is not in fact $\dot{y}$ but rather $\sign(\det(\mathbf{e})) \dot{y}$. In doing so we would see that we have swapped diads across the join. An equivalent way to see this is to consider that the conserved quantity corresponds to the interior product of the Hamiltonian vector field and the symplectic structure applied to the symmetry $s =\frac{\partial}{\partial y}$.  Upon parallel transporting $s_y$ across the join, it is inverted $s'_y=-\frac{\partial}{\partial y}$, and hence the conserved quantity takes the value $P'_y=-P_y$. However, perhaps the most intuitive way to understand this is to consider that the Lagrangian is equivalent to defining on the double cover 
\be \L' = \frac{\dot{x}^2}{2} + \frac{\dot{Y}^2}{2} \ee
and the conserved quantity really corresponds to $P_Y = \dot{Y}$. 

The issue of how chirality is treated in a relational context has been a long-standing issue in the philosophy of physics, dating back to Kant. The fundamental issue relates to objects which are counterparts, yet incongruous, with the common example being left and right hands. In the case we examine here, we find that within our system the descriptions are intrinsically indistinguishable; one a notion of handedness has been chosen, one can indeed distinguish between, clockwise and anticlockwise. However, given a single rotating object the system would not be sufficient to determine whether that object rotated clockwise or anticlockwise. For an excellent review of such issues see \cite{Pooley} and references therein\footnote{We are grateful to Sean Gryb for making pointing out the extant literature in response to an early draft of this paper.}. 

Let us now examine what happens in the FLRW case. Since the dynamics is captured in essence by the case of two massless scalar fields, we will take this as our example, however the results we obtain are more generic. We noted in section \ref{Free} that the solution $\phi_i = A_i \log(t) + B_i$ are continued across the singularity by extending to $t<0$ through taking $\phi_i = A_i \log |t| + \sign(t)B_i$, and thus $\dot{\phi}_i = \sign(t) \frac{A_i}{t}$. Naively we might think that this inversion had violated a conservation law; the reversal of a velocity would seem to be in contradiction to the conservation of momentum. However, recall that the momentum that is conserved in an FLRW cosmology is $p_\phi=v\dot{\phi}$. Thus, since in our solution $v \propto t$ we see that this is indeed a conserved charge if we make the stipulation that $v$ is the \textit{signed} volume of the fiducial cell. Equivalently, we should consider the connection across the singularity as being moving from the description of the spacetime geometry by the right-handed tetrad $\mathbf{e}_R$ to the left-handed triad $\mathbf{e}_L$. In this case the conservation law still holds; it is simply that in moving from one representation of our system to another it appears that we have undergone a parity inversion.

Solving the Friedmann equation and interpreting the evolution as being that of a space-time geometry away from the singularity we see that the metric can be expressed in the usual FLRW terms appearing as two universes joined back-to-back at the singularity. However, in order to have the conserved quantities remain conserved across the singularity itself, we must switch tetrads across this point. This appears, choosing to represent $s=\sign(t)$:
\be \d s^2 = -\d t^2 + |t|^{\frac{2}{3}} \left(\d x^2+\d y^2+\d z^2 \right) \quad \mathbf{e}= (\d t, s|t|^{\frac{1}{3}} (\d x,\d y,\d z)) \label{extended} \ee
and hence we can uniquely extend our space-time across the singularity in this manner.

At this point the question of the Hawking--Penrose singularity theorems \cite{Hawking} arises. In particular, how can we know that geodesics are connected across the singularity? One might expect that since the spatial metric has vanished at the singularity, any path orthogonal to the time direction will have no length, and thus we could connect any points in space here. In essence this is the infinite focussing on which the theorems are based. To answer this we must again engage with a relational description; consider the null geodesic to be the path taken by a photon on the space-time, ignoring backreaction. In coordinate terms, let us choose without loss of generality a null geodesic whose spatial component is parallel to the x axis in the region of space-time covered by $t>0$. The equation of motion is then, in coordinate terms, $x=x_\mathrm{0} + \frac{3}{2} t^\frac{2}{3}$. If we choose, as we did in the case of the Mobius strip example, to enforce continuity between points of the space-time manifold covered by different choices of tetrad, and further to conserve the momentum of the photon as measured in this way (respecting the signed volume of space), then there is a unique geodesic on the other side of the singularity which connects to this point. This is given by $x=x_o + \frac{3}{2} |t|^\frac{2}{3}$, with the tetrad given in equation (\ref{extended}).  We note that the conditions of the Hawking--Penrose theorem are satisfied, and hence the conclusion that there is infinite focusing of geodesics is valid. The distance between any two photons as measured using the space-time geometry will tend to zero as we approach the singularity. However, there is still a unique continuation of the paths beyond this point, which corresponds to the above trajectory as described in the geometry in which we swap tetrads. What the singularity theorems require is that we model physics in terms of a space-time defined by a metric on a globally orientable manifold. However, if we relax that assumption just on the singular surface itself, we see that there is a natural physical continuation beyond this point when described in relational terms. 

To be clear about this continuation, we note that the Hawking--Penrose theorems show that there is not a unique continuation through the singularity of the space-time geometry as determined by the Einstein equations. Our results are completely in agreement with this. We do not claim that the Einstein equations give such a continuation. However, the relational system of scalar fields can be continued beyond this point, and at all points away from the singularity their behaviour is consistent with that described by the Einstein equations coupled to the matter. Thus we know that on both sides of the singularity there is a consistent space-time picture. What we have then shown is that if we want to further identify trajectories of matter across the singularity such that conservation laws, specifically the conservation of momentum, hold then this in turn makes a unique identification. In this identification we can connect two FLRW cosmologies back-to-back at the singularity and continue geodesics from one into the other. This continuation is motivated by our example of the M\"obius strip, and in common with that example we see that orientation is inverted in the continuation. To relate back to the issue of handedness, given a "left hand" on one side of the singularity we see that it is continued to a hand on the other side. The hands that are congruous to it on one side remain congruous on the other. Since two incongruous "hands" that propagate across the singularity can be arbitrarily chosen to be called left or right, we cannot uniquely say whether a left hand on one side of the singularity remains a left hand on the other. We are free to choose whether to call this again a left hand or a right hand as the physical system makes no distinction between the two. However, if we wish to retain the conserved quantities both in magnitude and sign across this point, we see that this choice is equivalent to taking a left hand on one side of the singularity and calling the object that it propagates to on the other a right hand. 

\section{Beyond Isotropy: The Bianchi I Spacetime} \label{BeyondSec}

The results we have obtained in the case of the flat FLRW system can be extended to anisotropic cosmological solutions. These are of particular importance in the study of cosmological singularities as they are believed to capture the complete behaviour of an inhomogeneous system \cite{BKL,Berger,Ringstrom,AR}. More precisely, the BKL conjecture states that in the vicinity of a singularity, the dynamics becomes local, oscillatory and vacuum dominated, with the only matter of import being massless scalar fields. Thus if we know the behaviour of homogeneous cosmologies with scalar fields, we should capture dynamics in the neighborhood of generic singularities. The case of a single free field providing the matter content was covered in \cite{Through}, where the spacetime in consideration was Bianchi IX, though the results are general. In this section we will show that the results of this paper are entirely in line with those. Here we will present the case of a Bianchi I spacetime; the results are in fact more generic, but since the asymptotic motion of quiescent solutions is parallel to that obtained in the Bianchi I case this will be sufficient to demonstrate how the result holds. The Bianchi classification labels geometries by the commutativity classes of the three Killing vector fields that define homogeneity. These in turn contribute to the Ricci scalar, which in the 3+1 decomposition of GR acts as a potential for anisotropies \cite{Ellis,Uggla,Uggla2}. This scales as $v^{4/3}$, and hence will differ from the potentials of scalar fields, so we will need an extension of our relational ideas to describe the resulting shape systems in full generality \cite{Indistinguish}. Fortunately in the case where the Killing vectors all commute the Ricci scalar is exactly zero, and thus we can bypass this issue. This is the Bianchi I system described with metric:
\be \d s^2 = -\d t^2 +v^{2/3} \left(e^{\frac{q_1}{2}-\frac{q_2}{\sqrt{6}}} \d x^2 + e^{-\frac{q_1}{2}-\frac{q_2}{\sqrt{6}}} \d y^2 + e^{\sqrt{\frac{2}{3}} q_2} \d z^2 \right) \ee
and when written in these terms, the gravitational Lagrangian for the anisotropic space-time has the same form as that of an isotropic space-time with massless scalar fields:
\be \L = v\left(-\frac{2}{3} \frac{\dot{v}^2}{v^2} +\frac{\dot{q}_1^2}{2} +\frac{\dot{q}_2^2}{2} +\L_m \right ) \ee
Therefore we can continue this system beyond the singularity in exactly the same way as we did for the case of scalar fields, identifying $q_1$ and $q_2$ with $\phi_1$ and $\phi_2$, say, and numbering the rest of our fields $\phi_3$ to $\phi_{n+2}$. In the case of a single additional massless scalar field, we see that the continuation of geodesic motion on the 3-sphere corresponds exactly to the continuation that was shown in \cite{Through} for the Bianchi I case. 
 
 \section{Discussion} 
 \label{SecDiscussion}

Let us recapitulate the major results shown in this paper. We have seen that the relational motion of scalar fields in an FLRW cosmology can be described entirely in terms of the fields themselves, and without direct reference to the geometry on which their dynamics takes place. This is due to the dynamical similarity of the space of solutions. We compactified the space of fields on a sphere, as this most closely matches the way in which Misner variables were modelled in the anisotropic case. In this description the relational equations of motion are derived directly from a contact Hamiltonian and thus need never reference the space-time geometry in the first place. When the potentials for the fields are well-behaved, these equations satisfy the conditions of the Picard-Lindel\"of theorem at the point corresponding to the singularity of the original space-time, and thus have a unique, deterministic extension beyond this point, and we have shown how this continuation is manifested in terms of the fields themselves. We then take these relational obesrvables as being the fundamental ontological content of our theory. As such we are able to reconstruct a space-time on the other side of the singularity, and by choosing to enforce the conservation across the singularity of quantities that are conserved everywhere else, we arrive at a unique description of the geometry. This description corresponds to having two FLRW cosmologies glued back-to-back at their respective singularities with an orientation reversal between them. In terms of tetrads, we swap a left-handed tetrad, $\mathbf{e_1}$, for a right-handed one, $\mathbf{e_2}$, in crossing. 

We must reiterate here that the continuation that we have described is not a direct consequence of Einstein's equations for the geometrical degrees of freedom. The equations of motion for quantities such as the Hubble rate become non-deterministic at the singularity, and therefore Einstein's equations provide no unique continuation by themselves. However, the equations of motion for the relational observables that constitute our empirically accessible content do indeed remain deterministic. Therefore we can reconstruct a geometrical description from this relational content. As the system exhibits both dynamical similarity and has a choice of orientation of the tetrads that determine the geometry, there is a choice in how this reconstruction is done. This choice is no different than that which is available to any observer modelling the evolution of an FLRW cosmology from observations; the value of the scale factor at any given time is a free choice (typically picked to be unity today for efficiency of description). Likewise any given metric can be equally described by left and right-handed tetrads. By exploiting this freedom to choose, at the singularity itself, how we glue together solutions we are able to find a description in which conserved quantities retain their values across the singularity. However, this choice is not in itself unique; we could have chosen to invert the conservation laws, for example, at the price of geodesics being sharply 'kinked' at the singularity itself. The important thing to note is that this choice is a choice of the geometric framework which we use to describe the system, but does not affect the underlying relational dynamics. Any such choice would give equivalent results. The choice we have made is simply the one in which a consistent idea of momentum (for example) can be made. Physically it holds no more meaning than describing the motion of a particle in a given frame until a certain time and then switching frames. Consider describing a particle that falls past a window in terms of its height above the base of the window until it reaches this base, and thereafter in terms of the height above the ground. The description in terms of coordinates may be discontinuous between frames, but the underlying physics is unaffected. 

Throughout this paper we have worked with scalar fields as our matter content. The motivation for this is twofold; first they are the simplest form of matter to encode in a Lagrangian framework and thus are simple to translate into the contact geometry in which we model our system, and second massless scalar fields are the dominant form of matter on approach to the initial singularity. Further insight into the ways in which we continue solutions across the singularity may be gained by working with vector fields or pseudo-scalars. For vector matter it will be necessary to work with anisotropic solutions, as a homogeneous vector field has by necessity a preferred direction and thus would break the isotropy of the system. 

Thus far our attention has been largely focussed on cosmological solutions and their singularities. There is a natural parallel with black hole interiors, as spherically symmetric space-times having spatial slices $\R \times S^2$ are Kantowski-Sachs cosmological solutions. Therefore it is possible to extend work in the anisotropic case to include these solutions and investigate the singularities at the centers of black holes. Preliminary work indicates that there is a unique continuation through these points, however the complete solution is a work in progress. 

The existence of a classical continuation of our system beyond an initial singularity poses significant questions for quantum gravity. For long it has been assumed that the indeterminacy introduced by GR at singularities is good motivation to treat such points as places to look for the effects of quantum gravity. In the FLRW case arguments are made along the lines of dimensional analysis. The energy density of any matter present in a cosmological system must approach the Planck density, at which point it is argued that quantum gravity effects should become important. However, in the case of anisotropic cosmologies there are singularities even in the vacuum solutions. Even in the presence of matter, on approach to the singularity it is believed that the anisotropic geometric contributions to the evolution are dominant over matter contributions. Since there exists a classical continuation this motivation is lost; it may be that we have been looking for quantum gravity in the wrong places. 

\section*{Acknowledgements}

The author is grateful to Julian Barbour, Sean Gryb and Flavio Mercati for helpful discussion and comments.

\bibliographystyle{ieeetr}
\bibliography{FLRW+SF}

\begin{thebibliography}{10}

\bibitem{LIGO}
B.~P. Abbott {\em et~al.}, ``{GW170817: Observation of Gravitational Waves from
  a Binary Neutron Star Inspiral},'' {\em Phys. Rev. Lett.}, vol.~119, no.~16,
  p.~161101, 2017.

\bibitem{Strings1}
M.~Grana, ``{Flux compactifications in string theory: A Comprehensive
  review},'' {\em Phys. Rept.}, vol.~423, pp.~91--158, 2006.

\bibitem{Strings2}
L.~McAllister and E.~Silverstein, ``{String Cosmology: A Review},'' {\em Gen.
  Rel. Grav.}, vol.~40, pp.~565--605, 2008.

\bibitem{Strings3}
M.~B. Green and J.~H. Schwarz, ``{Anomaly Cancellation in Supersymmetric D=10
  Gauge Theory and Superstring Theory},'' {\em Phys. Lett.}, vol.~149B,
  pp.~117--122, 1984.

\bibitem{Strings4}
I.~Antoniadis, N.~Arkani-Hamed, S.~Dimopoulos, and G.~R. Dvali, ``{New
  dimensions at a millimeter to a Fermi and superstrings at a TeV},'' {\em
  Phys. Lett.}, vol.~B436, pp.~257--263, 1998.

\bibitem{LQG1}
A.~Ashtekar, M.~Reuter, and C.~Rovelli, ``{From General Relativity to Quantum
  Gravity},'' 2014.

\bibitem{LQG2}
A.~Ashtekar and P.~Singh, ``{Loop Quantum Cosmology: A Status Report},'' {\em
  Class. Quant. Grav.}, vol.~28, p.~213001, 2011.

\bibitem{LQC1}
P.~Singh, ``{Loop quantum cosmology and the fate of cosmological
  singularities},'' {\em Bull. Astron. Soc. India}, vol.~42, p.~121, 2014.

\bibitem{LQC2}
K.~Liegener and P.~Singh, ``{Gauge invariant bounce from quantum geometry},''
  2019.

\bibitem{CST1}
L.~Bombelli, J.~Lee, D.~Meyer, and R.~D. Sorkin, ``Space-time as a causal
  set,'' {\em Phys. Rev. Lett.}, vol.~59, pp.~521--524, Aug 1987.

\bibitem{CST2}
D.~P. Rideout and R.~D. Sorkin, ``Evidence for a continuum limit in causal set
  dynamics,'' {\em Phys. Rev. D}, vol.~63, p.~104011, Apr 2001.

\bibitem{GFT}
S.~Gielen, D.~Oriti, and L.~Sindoni, ``{Cosmology from Group Field Theory
  Formalism for Quantum Gravity},'' {\em Phys. Rev. Lett.}, vol.~111, no.~3,
  p.~031301, 2013.

\bibitem{Shapes1}
J.~Barbour, ``Dynamics of pure shape, relativity and the problem of time,''
  {\em Lect.Notes Phys.}, vol.~633, pp.~15--35, 2003.

\bibitem{Shapes2}
F.~Mercati, {\em Shape dynamics: Relativity and relationalism}.
\newblock Oxford University Press, 2018.

\bibitem{Shapes3}
H.~Gomes, S.~Gryb, and T.~Koslowski, ``{Einstein gravity as a 3D conformally
  invariant theory},'' {\em Class. Quant. Grav.}, vol.~28, p.~045005, 2011.

\bibitem{Shapes4}
H.~Gomes, ``Poincar{\'e} invariance and asymptotic flatness in shape
  dynamics,'' {\em Phys. Rev.}, vol.~D88, p.~024047, 12 2012.

\bibitem{Through}
T.~A. Koslowski, F.~Mercati, and D.~Sloan, ``{Through the big bang: Continuing
  Einstein's equations beyond a cosmological singularity},'' {\em Phys. Lett.},
  vol.~B778, pp.~339--343, 2018.

\bibitem{FlavNew}
F.~Mercati, ``{Through the Big Bang in inflationary cosmology},'' 2019.

\bibitem{DynSim}
D.~Sloan, ``{Dynamical Similarity},'' {\em Phys. Rev.}, vol.~D97, no.~12,
  p.~123541, 2018.

\bibitem{Bravetti}
A.~{Bravetti}, H.~{Cruz}, and D.~{Tapias}, ``{Contact Hamiltonian mechanics},''
  {\em Annals of Physics}, vol.~376, pp.~17--39, Jan 2017.

\bibitem{Leon}
M.~{Lainz Valc{\'a}zar} and M.~{de Le{\'o}n}, ``{Contact Hamiltonian
  Systems},'' {\em arXiv e-prints}, p.~arXiv:1811.03367, Nov 2018.

\bibitem{Bravetti2}
A.~{Bravetti} and D.~{Tapias}, ``{Liouville{\textquoteright}s theorem and the
  canonical measure for nonconservative systems from contact geometry},'' {\em
  Journal of Physics A Mathematical General}, vol.~48, p.~245001, Jun 2015.

\bibitem{Foster}
S.~Foster, ``{Scalar field cosmologies and the initial space-time
  singularity},'' {\em Class. Quant. Grav.}, vol.~15, pp.~3485--3504, 1998.

\bibitem{Pooley}
O.~Pooley, ``Handedness, parity violation, and the reality of space,'' 2001.

\bibitem{Hawking}
S.~W. Hawking and R.~Penrose, ``{The Singularities of gravitational collapse
  and cosmology},'' {\em Proc. Roy. Soc. Lond.}, vol.~A314, pp.~529--548, 1970.

\bibitem{BKL}
V.~a. Belinsky, I.~m. Khalatnikov, and E.~m. Lifshitz, ``{A General Solution of
  the Einstein Equations with a Time Singularity},'' {\em Adv. Phys.}, vol.~31,
  pp.~639--667, 1982.

\bibitem{Berger}
B.~K. Berger, D.~Garfinkle, J.~Isenberg, V.~Moncrief, and M.~Weaver, ``{The
  Singularity in generic gravitational collapse is space - like, local, and
  oscillatory},'' {\em Mod. Phys. Lett.}, vol.~A13, pp.~1565--1574, 1998.

\bibitem{Ringstrom}
H.~Ringstrom, ``{The Bianchi IX attractor},'' {\em Annales Henri Poincare},
  vol.~2, pp.~405--500, 2001.

\bibitem{AR}
L.~Andersson and A.~D. Rendall, ``{Quiescent cosmological singularities},''
  {\em Commun. Math. Phys.}, vol.~218, pp.~479--511, 2001.

\bibitem{Ellis}
G.~F.~R. Ellis and M.~A.~H. MacCallum, ``{A Class of homogeneous cosmological
  models},'' {\em Commun. Math. Phys.}, vol.~12, pp.~108--141, 1969.

\bibitem{Uggla}
J.~M. Heinzle and C.~Uggla, ``{Mixmaster: Fact and Belief},'' {\em Class.
  Quant. Grav.}, vol.~26, p.~075016, 2009.

\bibitem{Uggla2}
C.~Uggla, ``{The Nature of generic cosmological singularities},'' in {\em
  {Recent developments in theoretical and experimental general relativity,
  gravitation and relativistic field theories. Proceedings, 11th Marcel
  Grossmann Meeting, MG11, Berlin, Germany, July 23-29, 2006. Pt. A-C}},
  pp.~73--89, 2007.

\bibitem{Indistinguish}
D.~Sloan, ``{The Indistinguishability of Certain Lagrangian Theories (in
  preparation)},'' 2019.

\end{thebibliography}

\end{document}